\documentclass[aip,prapplied,reprint]{revtex4-2} 

\usepackage[T1]{fontenc}
\usepackage{times}

\usepackage{graphicx}
\usepackage{booktabs}
\usepackage{placeins}
\usepackage{amsfonts, amsmath,amssymb, bm, graphicx}
\usepackage{siunitx}
\DeclareSIUnit{\belmilliwatt}{Bm}
\DeclareSIUnit{\dBm}{\deci\belmilliwatt}
\usepackage{lipsum}
\usepackage[english]{babel}

\newcommand{\figref}[2]{\hyperref[#1]{\ref{#1}(#2)}}
\newcommand{\figrefsub}[3]{\hyperref[#1]{\ref{#1}(#2)#3}}

\usepackage{physics}
\usepackage{lipsum}
\usepackage{changes}

\usepackage[colorlinks=true,allcolors=blue]{hyperref}
\usepackage{footmisc}

\usepackage{upgreek}

\usepackage{soul}

\makeatletter
\let\ORIbbl@fixname\bbl@fixname
\def\bbl@fixname#1{%
  \@ifundefined{languagealias@\expandafter\string#1}
    {\ORIbbl@fixname#1}
    {\edef\languagename{\@nameuse{languagealias@#1}}}%
}
\newcommand{\definelanguagealias}[2]{%
  \@namedef{languagealias@#1}{#2}%
}
\makeatother

\definelanguagealias{en}{english}

\begin{document}

\title{Coherent control of Floquet-engineered magnon frequency combs}

\author{Christopher Heins}
\affiliation{Helmholtz-Zentrum Dresden--Rossendorf, Institut f\"ur Ionenstrahlphysik und Materialforschung, D-01328 Dresden, Germany}
\affiliation{Fakult\"at Physik, Technische Universit\"at Dresden, D-01062 Dresden, Germany}

\author{Amelie L. Fehrmann}
\affiliation{Helmholtz-Zentrum Dresden--Rossendorf, Institut f\"ur Ionenstrahlphysik und Materialforschung, D-01328 Dresden, Germany}
\affiliation{Fakult\"at Physik, Technische Universit\"at Dresden, D-01062 Dresden, Germany}

\author{Lukas K\"orber}
\affiliation{Helmholtz-Zentrum Dresden--Rossendorf, Institut f\"ur Ionenstrahlphysik und Materialforschung, D-01328 Dresden, Germany}
\affiliation{Fakult\"at Physik, Technische Universit\"at Dresden, D-01062 Dresden, Germany}
\affiliation{Radboud University, Institute of Molecules and Materials, Heyendaalseweg 135, 6525 AJ Nijmegen, The Netherlands}
\author{Joo-Von Kim}
\affiliation{Centre de Nanosciences et de Nanotechnologies, CNRS, Universit\'e Paris-Saclay, 91120 Palaiseau, France}

\author{Attila K\'akay}
\affiliation{Helmholtz-Zentrum Dresden--Rossendorf, Institut f\"ur Ionenstrahlphysik und Materialforschung, D-01328 Dresden, Germany}

\author{Jürgen Fassbender}
\affiliation{Helmholtz-Zentrum Dresden--Rossendorf, Institut f\"ur Ionenstrahlphysik und Materialforschung, D-01328 Dresden, Germany}
\affiliation{Fakult\"at Physik, Technische Universit\"at Dresden, D-01062 Dresden, Germany}

\author{Katrin Schultheiss}\email{k.schultheiss@hzdr.de}
\affiliation{Helmholtz-Zentrum Dresden--Rossendorf, Institut f\"ur Ionenstrahlphysik und Materialforschung, D-01328 Dresden, Germany}

\author{Helmut Schultheiss}\email{h.schultheiss@hzdr.de}
\affiliation{Helmholtz-Zentrum Dresden--Rossendorf, Institut f\"ur Ionenstrahlphysik und Materialforschung, D-01328 Dresden, Germany}

\date{\today}


\begin{abstract}
Frequency combs represent a hallmark of coherence emerging from nonlinear dynamics, where periodic driving organizes energy into a precisely spaced spectral structure. Extending this concept to collective excitations in solids such as magnons, the quanta of spin waves in magnetically ordered materials, offers a powerful route to control energy flow, coherence, and information processing in condensed matter systems. 
Here, we demonstrate deterministic control of Floquet-engineered magnon frequency combs in magnetic vortices using nanosecond voltage pulses. By tuning the pulse duration and timing, we control the nonlinear energy transfer between magnons and the vortex core, enabling the Floquet-engineered initiation or suppression of magnon frequency combs far below their spontaneous instability threshold. This pulse-programmable interaction allows the vortex to sustain magnon-driven auto-oscillation with high phase stability, or to revert to its static ground state on demand. Our results establish vortex-based magnetic systems as a robust solid-state platform for Floquet engineering, bridging nonlinear spin dynamics with frequency conversion and coherent spin-based quantum devices.
\end{abstract}

\maketitle


Frequency combs are among the most striking manifestations of coherence emerging from nonlinear dynamics. When a system is driven periodically, energy can organize into a discrete set of phase-locked spectral lines, forming a comb in frequency space. Originally developed in optics, where they revolutionized precision metrology and spectroscopy\cite{Udem1999, Reichert2000, Udem2002}, frequency combs are now being explored in a growing range of platforms from phononic\cite{Ganesan2017,He2025} and photonic resonators\cite{DelHaye2007,Zhang2025} to superconducting circuits\cite{Erickson2014,Wang2024} and magnetic materials\cite{Koerner2022,Wu2024}, promising new opportunities for information processing and hybrid quantum technologies. 

In magnetism, frequency combs arise from the nonlinear interactions between magnons. These interactions allow magnons to exchange energy and momentum through three- and four-magnon scattering, giving rise to frequency conversion\cite{Nikolaev2024,Bejarano2024,Lan2025}, bistability\cite{Wang2024_magnonic}, and coherent sideband generation\cite{Chai2022}. While such effects have been harnessed to generate magnon frequency combs (MFCs) in waveguides via stimulated magnon–magnon scattering \cite{hulaSpinwaveFrequencyCombs2022} and in yttrium iron garnet spheres through magneto-elastic coupling \cite{xuMagnonicFrequencyComb2023}, they typically rely on continuous driving above an instability threshold, which limits deterministic control and tunability. Moreover, these systems exploit magnon bands, where propagation losses and dispersion curvature lead to rapid dephasing.

\begin{figure*}[]
    \includegraphics{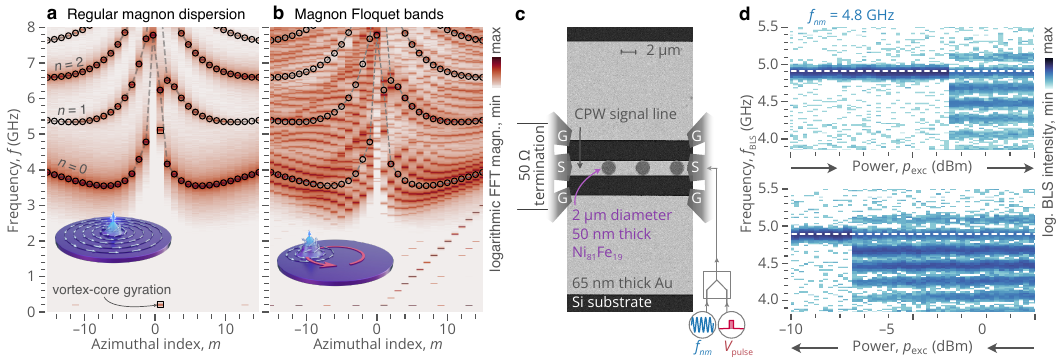}
    \caption{\textbf{From the regular magnon dispersion to Floquet-engineered frequency combs in a magnetic vortex.} \textbf{a}, Simulated regular magnon dispersion of a \SI{2}{\micro\meter}-diameter, \SI{50}{\nano\meter}-thick Ni$_{81}$Fe$_{19}$ disk in the vortex ground state, with the static magnetization configuration shown in the inset. High-frequency magnon modes characterized by their radial and azimuthal indices $(n, m)$ are marked by circles, while the low-frequency excitations corresponding to the thermally populated vortex-core gyration are indicated by squares.
    \textbf{b}, Floquet magnon dispersion under driven vortex-core gyration, as sketched in the inset. The circles overlay the regular magnon modes of the static vortex, highlighting the formation of additional Floquet sidebands and avoided crossings induced by the periodic drive. 
    \textbf{c}, Scanning electron microscopy image of Ni$_{81}$Fe$_{19}$ disks patterned on a gold coplanar waveguide (CPW). The excitation signal is provided through a diplexer combining the high-frequency microwave input ($f_{nm}$) with the nanosecond voltage pulses ($V_\text{pulse}$) applied via ground-signal-ground (GSG) probes. 
    \textbf{d}, Brillouin light scattering (BLS) microscopy spectra measured under single-tone excitation at $f_{nm} = \SI{4.8}{\giga\hertz}$, corresponding to the azimuthal mode $m = -1$, for varying excitation power. At sufficiently high powers, distinct magnon frequency combs emerge, marking the onset of nonlinear scattering processes and self-induced Floquet magnon states. The upper (lower) panel shows data obtained during power increase (decrease), revealing a clear hysteresis between the two sweep directions that reflects the threshold behaviour of the nonlinear magnon-magnon interaction.}
    \label{fig:geometry}
    \end{figure*}

Magnetic vortices offer an alternative platform for studying coherent nonlinear dynamics in confined geometries. In a vortex, the in-plane magnetization curls around a central core with an out-of-plane component (inset, Fig.~\ref{fig:geometry}a). Such textures support two distinct classes of excitations: GHz-frequency magnons and a MHz-frequency gyrotropic mode describing the orbiting motion of the core around its equilibrium position (inset, Fig.~\ref{fig:geometry}b). We have recently shown that the vortex-core gyration can act as an internal periodic drive, inducing Floquet magnon bands through nonlinear coupling between these modes~\cite{heins_self-induced_2024}. This dynamic modulation reshapes the regular magnon dispersion (Fig.~\ref{fig:geometry}a), introducing Floquet sidebands (Fig.~\ref{fig:geometry}b), which are accompanied by avoided level crossings and the generation of frequency combs. At strong driving amplitudes, even a single high-frequency magnon mode can self-induce Floquet magnons, spontaneously generating a frequency comb from a monochromatic excitation~\cite{heins_self-induced_2024,wang_twisted_2022}.

Here, we demonstrate the deterministic, time-resolved control of Floquet magnons and MFCs in magnetic vortices using nanosecond voltage pulses. By tuning the pulse duration and timing, we coherently steer the nonlinear energy exchange between high-frequency magnons and the vortex core, enabling on-demand generation and suppression of frequency combs far below their spontaneous instability threshold. In the active comb state, the vortex core becomes effectively undamped, allowing a self-sustained auto-oscillation driven purely by magnon torque. These results establish magnetic vortices as a model platform for time-domain control of nonlinear spin dynamics, opening new pathways toward reconfigurable frequency combs and illustrating a general route to time-domain control of nonlinear order in solids.

\section*{Results and discussion}

For our experiments, we patterned \SI{2}{\micro\meter}-diameter, \SI{50}{\nano\meter}-thick Ni$_{81}$Fe$_{19}$ disks on top of a coplanar waveguide (CPW), as shown in the scanning electron microscopy (SEM) image in Fig.~\figref{fig:geometry}{c}. The excitation combines a continuous GHz-frequency signal ($f_{nm}$) from a microwave generator with nanosecond voltage pulses ($V_\text{pulse}$) from an arbitrary waveform generator (AWG). Both signals are merged using a diplexer and applied through ground-signal-ground (GSG) probes. The associated in-plane magnetic field excites modes with $m=\pm1$. Depending on the excitation frequency, this  corresponds to the first azimuthal magnons or the vortex-core gyration. The resulting dynamics are detected by magneto-optical time-resolved Brillouin light scattering (BLS) microscopy (see Methods).

Figure~\ref{fig:geometry}d shows BLS spectra for $m=-1$ magnons excited at a fixed frequency of $f_{nm}=\SI{4.8}{\giga\hertz}$, while varying the microwave power between \SIrange[]{-10}{0}{\dBm}, going from low to high (top panel) and high to low values (bottom panel). Sweeping the power up to \SI{-3.5}{\dBm}, only the directly driven mode is visible. Above this threshold, the intensity of the main line decreases and an entire frequency comb emerges, with sidebands separated by the gyration frequency $f_\mathrm{gy}\approx\SI{200}{\mega\hertz}$. When the excitation power is subsequently reduced, the frequency comb persists down to \SI{-7.5}{\dBm}, revealing a pronounced hysteresis.

The ermergence of the MFC is governed by three-magnon splitting (3MS), which explains the abrupt transition between the single-tone and comb regimes. In this nonlinear process, a strongly excited high-frequency magnon decays into two lower-frequency magnons while conserving both energy and angular momentum, determined by the azimuthal index $m$. Here, one of the decay products corresponds to the vortex gyration, while the other occupies a mode with $f=f_{nm}-f_\text{gy}$ and $m=m_{nm}-m_\text{gy}$. This initial event launches a cascade of stimulated splitting and confluence processes, whereby populated magnons continue to split or merge, progressively filling lower and higher spectral orders and giving rise to the full MFC~\cite{korberNonlocalStimulationThreeMagnon2020,hula_spin-wave_2022,geNanoscaledMagnonTransistor2024}. When sweeping the excitation power downward, the existing vortex gyration lowers the effective instability threshold, sustaining the nonlinear scattering and producing the observed hysteresis. We now exploit this bistability to achieve deterministic switching between the static and dynamically modulated Floquet magnon dispersions.


\begin{figure*}[]
    \includegraphics{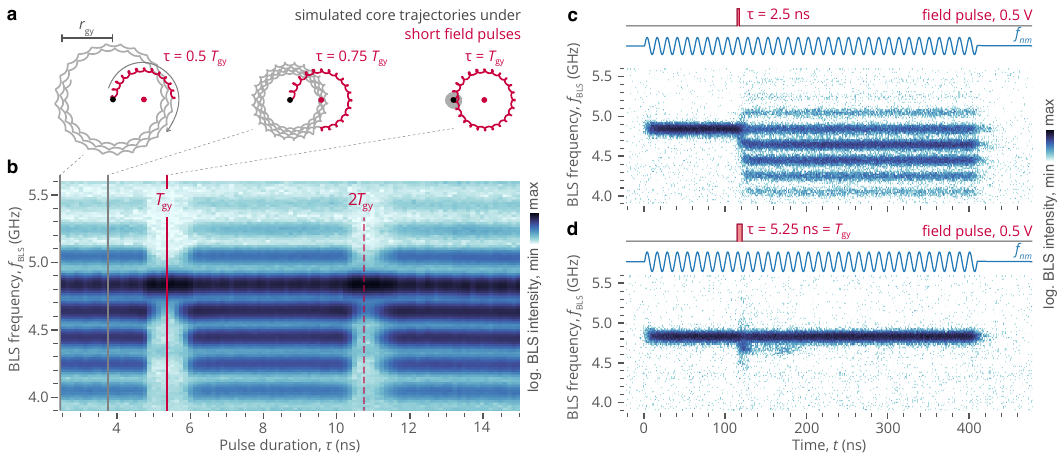}
    \caption{\textbf{Nanosecond-pulse control of Floquet-engineered magnon frequency combs in a magnetic vortex.}
    \textbf{a}, Vortex-core trajectories derived from micromagnetic simulations for different pulse durations $\tau$. Red (gray) traces denote the core motion during (after) the field pulse, and the red (black) circles mark the equilibrium positions during (after) excitation. A pulse with duration $\tau = 0.5T_\text{gy}$ displaces the core, triggering sustained gyration. In contrast, a pulse with $\tau = T_\text{gy}$ returns the core to the disk center at its falling edge, suppressing the gyration. 
    \textbf{b}, Time-integrated BLS spectra recorded as a function of pulse duration $\tau$. The Floquet states vanish whenever $\tau$ matches integer multiples of the vortex-core gyration period ($\tau = \ell T_\mathrm{gy}$, with $\ell = 1, 2, 3, …$), demonstrating that comb generation and suppression are governed by the phase relation between the external voltage pulses and the vortex-core motion.
    \textbf{c},\textbf{d}, Time-resolved BLS spectra measured on a Ni$_{81}$Fe$_{19}$ disk while exciting the $m = -1$ azimuthal mode at $f_{nm} = \SI{4.8}{\giga\hertz}$ with a \SI{400}{\nano\second} microwave pulse below the  threshold for spontaneous nonlinear coupling to the vortex-core gyration and self-induced Floquet states. At $t=\SI{120}{\nano\second}$, single voltage pulses of different duration are applied, as illustrated schematically above the spectra. A short pulse of $\tau = \SI{2.5}{\nano\second}$, close to half the gyration period, induces a clear magnon frequency comb, while a slightly longer pulse of $\tau = \SI{5.25}{\nano\second}$, approximately one gyration period, leaves the system in a single-tone state, suppressing the comb.
    }
    \label{fig:pulsed}
\end{figure*}

To this end, nanosecond voltage pulses are used to deflect the vortex core from its equilibrium position to initiate gyration, while simultaneously driving the $f_{nm}=\SI{4.8}{\giga\hertz}$ mode below its intrinsic instability threshold. To examine the core behavior first, we performed micromagnetic simulations using \textsc{mumax3} (see Methods). The simulated core trajectories are shown in Fig.~\ref{fig:pulsed}a. Before the pulse, the nonlinear coupling from the high-frequency magnons to the core is too weak to induce spontaneous gyration, and the core remains static at the disk center. With the onset of the voltage pulse, the associated magnetic field shifts the equilibrium position (from black to red dot in Fig.~\ref{fig:pulsed}a), initiating precession around the new equilibrium (red trace). The resulting trochoidal trajectory reflects the interplay between the gyration and high-frequency magnons. When the pulse ends after half a gyration period ($\tau = 0.5 T_\text{gy}$), the core motion continues on an outer orbit; for slightly longer pulses ($\tau = 0.75 T_\text{gy}$), the orbit shrinks. However, when the pulse ends after a full gyration ($\tau = T_\text{gy}$), the core returns to the center and remains static.

The simulated behavior is reminiscent of the oscillatory “ringing” observed during ultrafast magnetization reversal. There, a picosecond field pulse triggers precessional motion, and the final state depends sensitively on the pulse duration relative to the precession period~\cite{Back1998,Back1999,Bauer2000,Gerrits2002,Schumacher2002}. In a macrospin picture, a field pulse displaces the effective field $\mathbf{H}_\text{eff}$, causing the magnetization $\mathbf{M}$ to precess around it. When the pulse ends, $\mathbf{M}$ continues to precess if misaligned with its equilibrium field, or remains static if aligned. This non-inertial response (the instantaneous adaptation of $\mathbf{M}$ to field changes) is a direct analogue to the vortex-core dynamics observed here.

To demonstrate the impact of this behavior experimentally, a quasi-continuous microwave excitation ($f_{nm}=\SI{4.8}{\giga\hertz}$) is applied for \SI{400}{\nano\second}, and after \SI{120}{\nano\second} a \SI{0.5}{\volt} voltage pulse is superimposed with varying duration between \SIrange{2.5}{15}{\nano\second}. The time-integrated BLS spectra in Fig.~\ref{fig:pulsed}b confirm the simulated phase dependence as a function of the pulse duration. MFCs appear for nearly all pulse lengths except when $\tau=\ell T_\text{gy}$, with $\ell=1,2$, where the comb collapses entirely. For pulse durations approaching the gyration period, $\tau \approx T_\text{gy}$, the pulse leaves the core near the center, reducing the gyration radius and correspondingly the number of sidebands. This is consistent with the Floquet picture, where the number of sidebands scales with the gyration amplitude~\cite{heins_self-induced_2024}.

In two representative time-resolved BLS maps measured for $\tau = \SI{2.5}{\nano\second} \approx 0.5 T_\text{gy}$ and $\tau = \SI{5.25}{\nano\second} \approx T_\text{gy}$ (Figs.~\ref{fig:pulsed}c,d), only the drive frequency is visible at first, confirming operation below the instability threshold for self-induced Floquet generation. Upon arrival of the \SI{2.5}{\nano\second} pulse (Figs.~\ref{fig:pulsed}c), the spectral response changes abruptly: a frequency comb appears, with sidebands spaced by $f_\text{gy}$. The comb persists for the remainder of the microwave burst, demonstrating that the short pulse triggers sustained nonlinear coupling between the core and the driven magnon mode.

A qualitatively different behavior occurs when the pulse duration approximately matches the gyration period, $\tau = \SI{5.25}{\nano\second} \approx T_\text{gy}$ (Fig.~\ref{fig:pulsed}d). In this case, sidebands appear only transiently during the pulse and vanish immediately, restoring the single-frequency state. This phase-sensitive suppression confirms that the vortex-core position at the falling edge of the pulse determines whether energy is transferred to the nonlinear coupling channel.

    \begin{figure*}[]
    \includegraphics{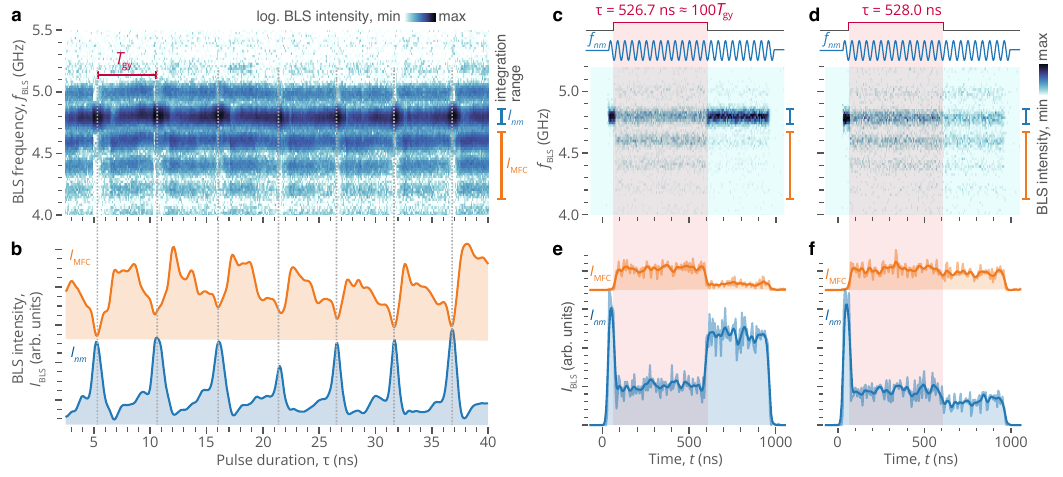}
    \caption{\textbf{Coherent control of magnon frequency-comb generation and suppression across all relevant timescales.}
    \textbf{a}, Time-integrated BLS spectra recorded as a function of pulse duration $\tau$. MFCs vanish whenever $\tau$ equals an integer multiple of the vortex-core gyration period ($\tau = \ell T_\text{gy}$, with $\ell = 1, 2, 3, …$), indicating that comb generation depends sensitively on the timing of the pulse within the gyration cycle. 
    \textbf{b}, BLS intensity integrated over the spectral ranges corresponding to the direct excitation ($I_{nm}$) and the lower-frequency comb modes ($I_\text{MFC}$), as sketched in panel (a). Peaks in $I_{nm}$ coincide with suppressed comb generation, reflecting that no energy is transferred into the sidebands when the pulse ends with the vortex core at the disk center.
    \textbf{c},\textbf{d}, Extended time-resolved BLS spectra for long pulses covering hundreds of gyration periods. In \textbf{c}, a \SI{526.7}{\nano\second} pulse ($\approx 100T_\text{gy}$) terminates when the vortex core is near the disk center, restoring the static configuration and suppressing the comb. In \textbf{d}, increasing the pulse length by only \SI{1.4}{\nano\second} leaves the core displaced at the falling edge, sustaining the gyration and preserving the frequency comb after the drive ends.
    \textbf{e},\textbf{f}, Temporal evolution of the integrated intensities for the direct excitation ($I_{nm}$) and lower-frequency comb sidebands ($I_\text{MFC}$). During the pulse, the direct excitation diminishes as energy is redistributed into the comb modes. When the pulse ends at the gyration period (d), the excitation recovers; when the pulse ends out of phase (f), the gyration remains, maintaining the comb.}
    \label{fig:extended_pulse}
    \end{figure*}

Through their interaction with the vortex core, excited Floquet modes influence the gyration dynamics by mitigating the effects of core damping. If only the nanosecond voltage pulses were considered, the vortex core would display similar trajectories but its motion would decay rapidly. In the present experiments, however, high-frequency magnons are continuously driven and scatter into the gyration, thereby replenishing the energy lost to damping. This sustained response occurs below the spontaneous instability threshold for two reasons. 
First, as shown in Figs.~\ref{fig:geometry}a,b, activation of the core modifies the magnon dispersion, introducing frequency shifts and additional sidebands. These modifications open new scattering channels and alter the three-magnon scattering coefficients, effectively lowering the threshold required to sustain auto-oscillation.
Second, once the core motion is established, stimulated scattering favors decay into already populated modes~\cite{korberNonlocalStimulationThreeMagnon2020,hula_spin-wave_2022,geNanoscaledMagnonTransistor2024}, further enhancing the probability that excited magnons feed the gyration. Theory and simulations confirm that multiple stable gyration radii can be maintained through this nonlinear interaction~\cite{Gauthier2025}.

Extending the pulse duration over several gyration periods demonstrates that the coupling between high-frequency magnons and the vortex core is coherent (Fig.~\ref{fig:extended_pulse}a). Even in stroboscopically acquired data, the suppression effect persists over many gyration periods. The integrated BLS intensities in Fig.~\ref{fig:extended_pulse}b are evaluated around the direct excitation ($I_{nm}$) and the lower sidebands of the MFC ($I_\text{MFC}$), and reveal a clear phase dependence: whenever the pulse duration equals an integer multiple of the gyration period ($\tau = \ell T_\text{gy}$, vertical dashed lines in Fig.~\ref{fig:extended_pulse}b), the intensity of the direct excitation peaks while the sideband intensity diminishes. This behavior indicates that the vortex trajectories remain phase-coherent and well defined over many successive periods, enabling deterministic control over extended timescales.

To explore this long-term stability, we increased the pulse duration up to one hundred gyration periods. Two representative cases are displayed in Figs.~\ref{fig:extended_pulse}c,d: one in which the pulse length is an integer multiple of $T_\text{gy}$ and one that ends with the core displaced from the center. The corresponding time-resolved BLS maps reveal complementary behavior. For the integer-multiple pulse ($\tau=\SI{526.7}{\nano\second}\approx 100 T_\text{gy}$), the direct excitation regains intensity upon pulse termination, while the comb sidebands collapse. When the pulse exceeds $100T_\text{gy}$ by \SI{1.4}{\nano\second}, the comb persists and the direct excitation remains weak.

To quantify the coupling between the high-frequency mode and the comb sidebands, the BLS intensity was integrated over the direct excitation ($I_{nm}$) and the lower-order sidebands ($I_\text{MFC}$) (Figs.~\ref{fig:extended_pulse}{e,f}). 
Both signals exhibit changes as a function of time, with minima in $I_{nm}$ coinciding with maxima in $I_\text{MFC}$. 
This coherent, energy-conserving transfer between the driven and comb modes confirms that the nonlinear coupling is phase-locked to the vortex motion, enabling deterministic control of the MFC through pulse timing.

    \begin{figure}
    \includegraphics{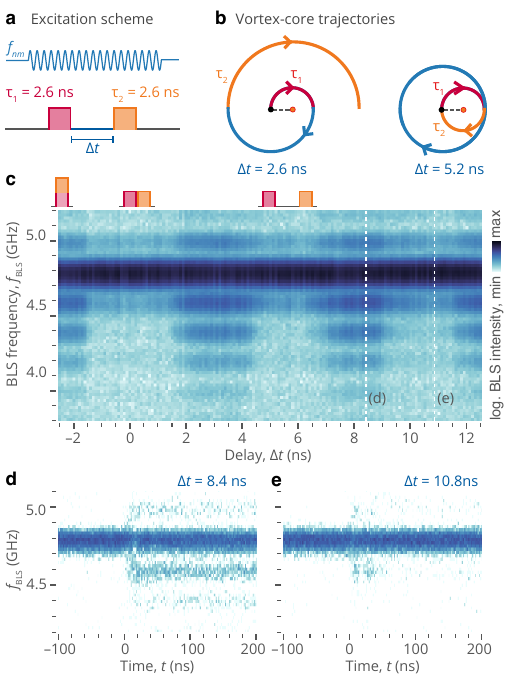}
    \caption{\textbf{Phase-controlled switching of magnon frequency combs using double-pulse excitation.} \textbf{a}, Modified excitation scheme with two voltage pulses of equal duration, $\tau_1=\tau_2=\SI{2.6}{\nano\second}$, separated by a delay $\Delta t$. 
    \textbf{b}, Schematic vortex-core trajectories for two representative delays. In both cases, the first pulse ($\tau_1$) shifts the equilibrium position from the black dot to the red dot, initiating the gyration over half a period (red trace). For $\Delta t =\SI{2.6}{\nano\second}$, the core completes half a precession around the disk center (blue trace) before the second pulse ($\tau_2$) arrives, expanding the orbit further (yellow trace). For $\Delta t =\SI{5.2}{\nano\second}$, corresponding to a full gyration period, the second pulse drives the core back to the disk center, quenching the gyration.
    \textbf{c}, Time-integrated BLS spectra as a function of pulse delay $\Delta t$. Whenever the delay equals in integer multiple of the gyration period, $\Delta t = \ell T_\text{gy}$, with $\ell =1,2,3...$, the comb sidebands and Floquet features are suppressed. 
    \textbf{d}, \textbf{e}, Time-resolved BLS spectra for delays $\Delta t=\SI{8.4}{\nano\second}$ and \SI{10.8}{\nano\second}, corresponding to $1.6 T_\text{gy}$ and $2 T_\text{gy}$ respectively. For the shorter delay, the gyration is sustained and a MFC forms; for the longer delay, the motion is suppressed and only single-frequency excitation remains. }
    \label{fig:double_pulse}
    \end{figure}

To demonstrate active toggle switching of the MFC, we replaced the long excitation pulse with two short, time-separated pulses ($\tau_1=\tau_2=\SI{2.6}{\nano\second}$), separated by variable delay $\Delta t$, as illustrated in Fig.~\ref{fig:double_pulse}a.  Figure~\ref{fig:double_pulse}b sketches the corresponding vortex-core trajectories for two representative delays. The first pulse displaces the core from the center and drives gyration over half a period around the shifted equilibrium position (red trace). During the delay, the core continues to orbit around the disk center through energy transfer from the high-frequency magnons (blue trace). The second pulse (yellow trace) then either amplifies the motion, expanding the orbit, or quenches it, depending on the phase of the ongoing gyration. Suppression occurs when the delay coincides with an integer multiple of the gyration period, $\Delta t = \ell T_\text{gy}$, with $\ell=1,2,3,\dots$.  

The experimental results in Fig.~\ref{fig:double_pulse}c confirm this scenario. For overlapping pulses ($\Delta t=\SI{-2.5}{\nano\second}$), displacement is enhanced and a MFC emerges. For $\Delta t=\SI{0}{\nano\second}$, equivalent to a combined pulse duration of one gyration period, the core motion halts completely and only the single-frequency excitation remains. At finite delays, the double-pulse excitation produces the expected periodic modulation of the MFC intensity: the comb vanishes whenever $\Delta t=\ell T_\text{gy}$ and reappears for intermediate values.
Time-resolved BLS spectra for $\Delta t=\SI{8.4}{\nano\second}$ and $\Delta t=\SI{10.8}{\nano\second}\approx 2T_\text{gy}$ (Figs.~\ref{fig:double_pulse}d,e) illustrate this phase-controlled on/off switching of Floquet-magnon generation.


\section*{Conclusion}

We have demonstrated that the hysteretic nonlinear coupling between GHz-frequency magnons and the vortex gyration provides a powerful means to switch between undriven and dynamically modulated magnon dispersions on demand. Using short voltage pulses, Floquet states can be excited well below their intrinsic instability threshold, with the outcome governed by the pulse duration: suitably timed pulses either confine magnon frequency combs to the pulse window or sustain them after the pulse ends. The strong nonlinear interaction enables GHz-frequency magnons to compensate damping of the vortex gyration, supporting persistent oscillations once initiated. Moreover, the coupling between magnon modes and the vortex core remains highly coherent, allowing deterministic switching even under extended pulse excitation, where the gyration can be quenched precisely at the falling edge.

Building on this, we have shown that two temporally separated pulses introduce an additional control dimension, enabling phase-sensitive toggle switching of the Floquet dispersion. Compared with a continuous drive, the double-pulse protocol achieves full control while reducing the required energy input, combining efficiency with enhanced selectivity of the vortex core excitation.

These findings reveal a fundamental aspect of self-induced Floquet dynamics in magnetic systems: coherent, phase-controlled manipulation of nonlinear excitations through pulsed driving. 
Beyond magnetism, this mechanism exemplifies a universal strategy for time-domain control in nonlinear wave systems. In the broader context of spin-based information processing, the demonstrated pulse-programmable generation and suppression of magnon frequency combs opens a pathway toward reconfigurable, energy-efficient spintronic oscillators and neuromorphic architectures driven by tailored nonlinear dynamics.
At the same time, they establish a practical route toward on-demand generation and suppression of magnon frequency combs, offering a flexible platform for magnonic frequency combs and future spintronic devices.

\FloatBarrier


\bibliographystyle{apsrev4-2}
\bibliography{references.bib}

\section*{Methods}

\subsection*{Sample preparation}
The sample was fabricated in a two-step procedure. We started with an undoped, high-resistance silicon substrate. In a first step, we patterned the coplanar waveguide (CPW) using a double-layer resist of methyl methacrylate (EL11) and poly(methyl methacrylate) (950 PMMA-A2), electron beam lithography, electron beam evaporation of a Cr(5~nm)/Au(65~nm) layer and subsequent lift-off in an acetone bath. The central line and ground lines of the CPW have widths of \SI{2}{\micro\meter} and \SI{13.5}{\micro\meter}, respectively, the gap between them is \SI{2.8}{\micro\meter} wide.

In a second step, the magnetic structures are patterned directly on top of the central signal line of the CPW. Therefore, we use a poly(methyl methacrylate) (950 PMMA-A6) resist, electron beam lithography, electron beam evaporation of a Cr(5~nm)/Ni$_{81}$Fe$_{19}$(50~nm)/Cr(2~nm) layer and subsequent lift-off in an acetone bath.

\subsection*{Time-resolved Brillouin light scattering microscopy}

The high-frequency magnons are detected by Brillouin light scattering (BLS) microscopy~\cite{sebastian_micro-focused_2015}. Therefore, a monochromatic, continuous-wave \SI{532}{\nano\meter} laser was focused onto the sample surface using a microscope objective with a high numerical aperture. The backscattered light was then directed into a Tandem Fabry-P\'{e}rot interferometer~\cite{mock_construction_1987} in order to measure the frequency shift caused by the inelastic scattering of photons and magnons. The detected intensity of the frequency-shifted signal is directly proportional to the magnon intensity at the respective focusing position. 

The response in the time-domain is probed by the time-of-flight principle. Therefore, we simultaneously monitor the state of the interferometer and the time when each photon is detected with respect to a clock provided by the microwave generator using a time-to-digital converter. In order to acquire enough signal, the pulsed experiment needs to be repeated stroboscopically, covering hundred thousands of repetitions. 

During all experiments, the investigated  microstructure is imaged using a red LED and a CCD camera. Displacements and drifts of the sample were tracked by an image recognition algorithm and compensated by the sample positioning system.

The magnetic response is measured at a single position on the disk, away from the vortex center. Because the sidebands of the magnon frequency comb correspond to propagating modes, they can all be detected at this location. The laser position is kept fixed, as local heating can modify the threshold for nonlinear processes. For instance, it has been shown that thermal gradients can displace the vortex core~\cite{Vogel2023}. Here, we aim to induce the transition using a fixed scheme. Therefore, it is crucial that the threshold is not altered by laser heating.

\subsection*{Electrical setup}

The quasi-continuous microwave excitation is provided by a signal generator. To ensure an excitation below the spontaneous coupling to the gyration, before each measurement a power sweep was conducted and the excitation power for the following measurements were chosen to stay \SI{1.5}{\dBm} below the threshold. The field pulses were applied by an arbitrary waveform generator (Keysight 81160A). In Fig.~\ref{fig:pulsed} the amplitude was \SI{0.5}{\volt}, in Fig.~\ref{fig:extended_pulse} the amplitude was increased to \SI{1}{\volt}, and in Fig.~\ref{fig:double_pulse} to \SI{0.8}{\volt} to ensure the gyration trajectory intersects with the disk center.

\subsection*{Micromagnetic simulations}

Simulations of the vortex dynamics were performed using the open-source finite-difference micromagnetics package \textsc{MuMax3}~\cite{vansteenkiste_design_2014}. 
The NiFe disk, with a diameter of \SI{2}{\micro\meter} and a thickness of \SI{50}{\nano\meter}, was discretized into a $512\times512\times1$ mesh. 
The material parameters used were the gyromagnetic ratio $\gamma = 1.76\times10^{11}\,\si{\radian\per(\tesla\second)}$, 
saturation magnetization $M_\mathrm{s} = 796~\si{kA/m}$, exchange stiffness $A_\mathrm{ex} = 13~\si{pJ/m}$, 
and Gilbert damping constant $\alpha = 0.007$.

A continuous sinusoidal in-plane excitation field,
\begin{equation*}
    h_\mathrm{exc} = b_\mathrm{exc}\sin(2\pi f_\mathrm{exc} t)\,\hat{\mathbf{y}},
\end{equation*}

with amplitude $b_\mathrm{exc} = \SI{0.15}{\milli\tesla}$ and frequency $f_\mathrm{exc} = \SI{4.8}{\giga\hertz}$, 
was applied to excite spin waves below the onset of spontaneous coupling to vortex gyration. 
At \SI{20}{\nano\second}, a short in-plane field pulse,
\begin{equation*}
b_\mathrm{pulse} = \SI{1.5}{\milli\tesla}\,\hat{\mathbf{x}},
\end{equation*}
was applied to displace the vortex core from its equilibrium position and initiate the gyration. 
The pulse duration was chosen to correspond to $0.5T_\mathrm{gy}=\SI{2.5}{\nano\second}$, $0.75T_\mathrm{gy}=\SI{3.75}{\nano\second}$, and $1T_\mathrm{gy}=\SI{5}{\nano\second}$ respectively. 
Throughout the simulation, the position of the vortex core was determined by tracking the cell with the maximum out-of-plane magnetization.

The magnon dispersion relations in Figs.~\figref{fig:geometry}{a,b} were simulated as described in Ref.~\citenum{heins_self-induced_2024}.

\section*{Data availability}

The data that support the findings of this study are openly available in RODARE\cite{heins_christopher_2025_4105}. 

\section*{Acknowledgments}

The authors thank B. Scheumann for depositing the Ni$_{81}$Fe$_{19}$ and Au films. Support by the Nanofabrication Facilities Rossendorf (NanoFaRo) at the IBC is gratefully acknowledged. The project has received funding by the EU Research and Innovation Programme Horizon Europe under grant agreement no. 101070290 (NIMFEIA) (C.H., K.S.). L.K. gratefully acknowledges funding by the Radboud Excellence Initiative.

\subsection*{Author contributions}

Conceptualization: H.S., C.H., K.S.
Investigation: C.H., A.F.
Data analysis: C.H., A.F.
Micromagnetic simulations: C.H., J.-V.K.
Sample fabrication: K.S.
Visualization: C.H., L.K., H.S., K.S.
Funding acquisition: H.S., K.S., J.-V.K.
Project administration: H.S., K.S.
Writing – original draft: C.H., K.S.
Writing – review and editing: all authors

\subsection*{Competing interests}
The authors declare no competing interests.

\end{document}